\documentclass[12pt]{article}

\newcounter{lastnote}

\usepackage{amssymb,amsmath,graphicx,times,epic,eepic,jtb}

\usepackage{cite}

\begin{document}

\title{Epidemic outbreaks on structured populations}

\author{Alexei Vazquez\\
The Simons Center for Systems Biology\\
Institute for Advanced Study, Einstein Dr, Princeton, NJ 08540, USA}

\date{\today}

\maketitle

\begin{abstract}

Our chances to halt epidemic outbreaks rely on how accurately we represent
the population structure underlying the disease spread.  When analyzing
global epidemics this force us to consider metapopulation models taking
into account intra- and inter-community interactions. Recently Watts {\it
et al} introduced a metapopulation model which accounts for several
features observed in real outbreaks [Watts {\it et al}, PNAS 102, 11157
(2005)]. In this work I provide an analytical solution to this model,
enhancing our understanding of the model and the epidemic outbreaks it
represents. First, I demonstrate that depending on the intra-community
expected outbreak size and the fraction of social bridges the epidemic
outbreaks die out or there is a finite probability to observe a global
epidemics. Second, I show that the global scenario is characterized by
resurgent epidemics, their number increasing with increasing the
intra-community average distance between individuals.  Finally, I present
empirical data for the AIDS epidemics supporting the model predictions.

\end{abstract}

\bibliographystyle{jtb}

Human populations are structured in communities representing geographical
locations and other factors leading to partial segregation. This
population structure has a strong impact on the spreading patterns of
infectious diseases among humans, forcing us to consider metapopulation
models making an explicit distinction between the intra- inter-community
interactions \cite{rvachev85,sattenspiel95}. The increase in model realism
is paid, however, by an increase in model complexity. Detailed
metapopulation models are difficult to build and as a consequence they are
available for a few locations in the world
\cite{rvachev85,flahault88,eubank04,germann06} or they cover a single
route of global transmission \cite{hufnagel04,colizza06}.

Recently Watts {\it et al} \cite{watts05} introduced a simple 
metapopulation model making an explicit distinction between the intra- and 
inter-community interactions. In spite of the model simplicity it accounts 
for several features observed in real epidemic outbreaks. In particular, 
the numerical results indicate the existence of a transition from local to 
global epidemics when the expected number of infected individuals changing 
community reaches one \cite{watts05}. I go a step forward and provide an 
analytical solution to the Watts {\it et al} metapopulation model. I 
demonstrate that there is indeed a phase transition when the expected 
number of infected individuals changing community reaches one. This 
analytical solution allow us to obtain a much deeper insight into the main 
features of global epidemic outbreaks.

\begin{figure}

\begin{center}
\setlength{\unitlength}{0.00041667in}
\begingroup\makeatletter\ifx\SetFigFont\undefined%
\gdef\SetFigFont#1#2#3#4#5{%
  \reset@font\fontsize{#1}{#2pt}%
  \fontfamily{#3}\fontseries{#4}\fontshape{#5}%
  \selectfont}%
\fi\endgroup%
{\renewcommand{\dashlinestretch}{30}
\begin{picture}(10226,10535)(0,-10)
\thicklines
\put(8713,2708){\ellipse{2996}{2996}}
\path(835,9012)(1435,8712)
\path(1193.505,8765.666)(1435.000,8712.000)(1247.170,8872.997)
\path(1435,8712)(2035,8412)
\path(1793.505,8465.666)(2035.000,8412.000)(1847.170,8572.997)
\path(1435,8712)(2035,9012)
\path(1847.170,8851.003)(2035.000,9012.000)(1793.505,8958.334)
\path(835,9012)(1435,9312)
\path(1247.170,9151.003)(1435.000,9312.000)(1193.505,9258.334)
\path(2035,9012)(2635,9012)
\path(2395.000,8952.000)(2635.000,9012.000)(2395.000,9072.000)
\path(2035,9012)(2635,8712)
\path(2393.505,8765.666)(2635.000,8712.000)(2447.170,8872.997)
\path(2635,9012)(3235,9012)
\path(2995.000,8952.000)(3235.000,9012.000)(2995.000,9072.000)
\path(1435,9312)(2035,9312)
\path(1795.000,9252.000)(2035.000,9312.000)(1795.000,9372.000)
\path(2035,9312)(2635,9312)
\path(2395.000,9252.000)(2635.000,9312.000)(2395.000,9372.000)
\path(2035,9312)(2635,9612)
\path(2447.170,9451.003)(2635.000,9612.000)(2393.505,9558.334)
\path(2635,9312)(3235,9312)
\path(2995.000,9252.000)(3235.000,9312.000)(2995.000,9372.000)
\path(1435,9312)(2035,9612)
\path(1847.170,9451.003)(2035.000,9612.000)(1793.505,9558.334)
\path(2635,8712)(3235,8712)
\path(2995.000,8652.000)(3235.000,8712.000)(2995.000,8772.000)
\path(2635,8712)(3235,8412)
\path(2993.505,8465.666)(3235.000,8412.000)(3047.170,8572.997)
\dashline{120.000}(3235,9012)(4435,9012)
\blacken\path(4075.000,8922.000)(4435.000,9012.000)(4075.000,9102.000)(4183.000,9012.000)(4075.000,8922.000)
\dashline{120.000}(2035,8412)(235,5712)
\blacken\path(359.808,6061.461)(235.000,5712.000)(509.577,5961.615)(374.784,5921.677)(359.808,6061.461)
\dashline{120.000}(2635,5712)(3835,5412)
\blacken\path(3463.920,5412.000)(3835.000,5412.000)(3507.577,5586.626)(3590.524,5473.119)(3463.920,5412.000)
\dashline{120.000}(2035,4812)(535,2262)
\blacken\path(639.953,2617.928)(535.000,2262.000)(795.101,2526.665)(662.769,2479.207)(639.953,2617.928)
\dashline{120.000}(5635,4812)(7435,2712)
\blacken\path(7132.382,2926.761)(7435.000,2712.000)(7269.048,3043.904)(7271.001,2903.333)(7132.382,2926.761)
\path(235,5712)(835,5412)
\path(593.505,5465.666)(835.000,5412.000)(647.170,5572.997)
\path(835,5412)(1435,5112)
\path(1193.505,5165.666)(1435.000,5112.000)(1247.170,5272.997)
\path(835,5412)(1435,5712)
\path(1247.170,5551.003)(1435.000,5712.000)(1193.505,5658.334)
\path(235,5712)(835,6012)
\path(647.170,5851.003)(835.000,6012.000)(593.505,5958.334)
\path(1435,5712)(2035,5712)
\path(1795.000,5652.000)(2035.000,5712.000)(1795.000,5772.000)
\path(1435,5712)(2035,5412)
\path(1793.505,5465.666)(2035.000,5412.000)(1847.170,5572.997)
\path(2035,5712)(2635,5712)
\path(2395.000,5652.000)(2635.000,5712.000)(2395.000,5772.000)
\path(835,6012)(1435,6012)
\path(1195.000,5952.000)(1435.000,6012.000)(1195.000,6072.000)
\path(1435,6012)(2035,6012)
\path(1795.000,5952.000)(2035.000,6012.000)(1795.000,6072.000)
\path(1435,6012)(2035,6312)
\path(1847.170,6151.003)(2035.000,6312.000)(1793.505,6258.334)
\path(2035,6012)(2635,6012)
\path(2395.000,5952.000)(2635.000,6012.000)(2395.000,6072.000)
\path(835,6012)(1435,6312)
\path(1247.170,6151.003)(1435.000,6312.000)(1193.505,6258.334)
\path(2035,5412)(2635,5412)
\path(2395.000,5352.000)(2635.000,5412.000)(2395.000,5472.000)
\path(2035,5412)(2635,5112)
\path(2393.505,5165.666)(2635.000,5112.000)(2447.170,5272.997)
\path(1435,5112)(2035,5112)
\path(1795.000,5052.000)(2035.000,5112.000)(1795.000,5172.000)
\path(1435,5112)(2035,4812)
\path(1793.505,4865.666)(2035.000,4812.000)(1847.170,4972.997)
\path(535,2262)(1135,1962)
\path(893.505,2015.666)(1135.000,1962.000)(947.170,2122.997)
\path(1135,1962)(1735,1662)
\path(1493.505,1715.666)(1735.000,1662.000)(1547.170,1822.997)
\path(1135,1962)(1735,2262)
\path(1547.170,2101.003)(1735.000,2262.000)(1493.505,2208.334)
\path(535,2262)(1135,2562)
\path(947.170,2401.003)(1135.000,2562.000)(893.505,2508.334)
\path(1735,2262)(2335,2262)
\path(2095.000,2202.000)(2335.000,2262.000)(2095.000,2322.000)
\path(1135,2562)(1735,2562)
\path(1495.000,2502.000)(1735.000,2562.000)(1495.000,2622.000)
\path(1735,2562)(2335,2562)
\path(2095.000,2502.000)(2335.000,2562.000)(2095.000,2622.000)
\path(1735,2562)(2335,2862)
\path(2147.170,2701.003)(2335.000,2862.000)(2093.505,2808.334)
\path(2335,2562)(2935,2562)
\path(2695.000,2502.000)(2935.000,2562.000)(2695.000,2622.000)
\path(1135,2562)(1735,2862)
\path(1547.170,2701.003)(1735.000,2862.000)(1493.505,2808.334)
\path(1735,1662)(2335,1662)
\path(2095.000,1602.000)(2335.000,1662.000)(2095.000,1722.000)
\path(1735,1662)(2335,1362)
\path(2093.505,1415.666)(2335.000,1362.000)(2147.170,1522.997)
\path(3835,5412)(4435,5112)
\path(4193.505,5165.666)(4435.000,5112.000)(4247.170,5272.997)
\path(4435,5112)(5035,4812)
\path(4793.505,4865.666)(5035.000,4812.000)(4847.170,4972.997)
\path(3835,5412)(4435,5712)
\path(4247.170,5551.003)(4435.000,5712.000)(4193.505,5658.334)
\path(5635,5412)(6235,5412)
\path(5995.000,5352.000)(6235.000,5412.000)(5995.000,5472.000)
\path(4435,5712)(5035,5712)
\path(4795.000,5652.000)(5035.000,5712.000)(4795.000,5772.000)
\path(5035,5712)(5635,5712)
\path(5395.000,5652.000)(5635.000,5712.000)(5395.000,5772.000)
\path(5035,5712)(5635,6012)
\path(5447.170,5851.003)(5635.000,6012.000)(5393.505,5958.334)
\path(5635,5712)(6235,5712)
\path(5995.000,5652.000)(6235.000,5712.000)(5995.000,5772.000)
\path(4435,5712)(5035,6012)
\path(4847.170,5851.003)(5035.000,6012.000)(4793.505,5958.334)
\path(5635,5112)(6235,5112)
\path(5995.000,5052.000)(6235.000,5112.000)(5995.000,5172.000)
\path(5635,5112)(6235,4812)
\path(5993.505,4865.666)(6235.000,4812.000)(6047.170,4972.997)
\path(5035,4812)(5635,4812)
\path(5395.000,4752.000)(5635.000,4812.000)(5395.000,4872.000)
\path(5035,4812)(5635,4512)
\path(5393.505,4565.666)(5635.000,4512.000)(5447.170,4672.997)
\path(4435,9012)(5035,8712)
\path(4793.505,8765.666)(5035.000,8712.000)(4847.170,8872.997)
\path(5035,8712)(5635,8412)
\path(5393.505,8465.666)(5635.000,8412.000)(5447.170,8572.997)
\path(5035,8712)(5635,9012)
\path(5447.170,8851.003)(5635.000,9012.000)(5393.505,8958.334)
\path(4435,9012)(5035,9312)
\path(4847.170,9151.003)(5035.000,9312.000)(4793.505,9258.334)
\path(5635,9012)(6235,9012)
\path(5995.000,8952.000)(6235.000,9012.000)(5995.000,9072.000)
\path(6235,9012)(6835,9012)
\path(6595.000,8952.000)(6835.000,9012.000)(6595.000,9072.000)
\path(5035,9312)(5635,9312)
\path(5395.000,9252.000)(5635.000,9312.000)(5395.000,9372.000)
\path(5635,9312)(6235,9312)
\path(5995.000,9252.000)(6235.000,9312.000)(5995.000,9372.000)
\path(5635,9312)(6235,9612)
\path(6047.170,9451.003)(6235.000,9612.000)(5993.505,9558.334)
\path(6235,9312)(6835,9312)
\path(6595.000,9252.000)(6835.000,9312.000)(6595.000,9372.000)
\path(5035,9312)(5635,9612)
\path(5447.170,9451.003)(5635.000,9612.000)(5393.505,9558.334)
\path(5635,8412)(6235,8412)
\path(5995.000,8352.000)(6235.000,8412.000)(5995.000,8472.000)
\path(5635,8412)(6235,8112)
\path(5993.505,8165.666)(6235.000,8112.000)(6047.170,8272.997)
\path(7435,2712)(8035,2412)
\path(7793.505,2465.666)(8035.000,2412.000)(7847.170,2572.997)
\path(8035,2412)(8635,2112)
\path(8393.505,2165.666)(8635.000,2112.000)(8447.170,2272.997)
\path(7435,2712)(8035,3012)
\path(7847.170,2851.003)(8035.000,3012.000)(7793.505,2958.334)
\path(8035,3012)(8635,3012)
\path(8395.000,2952.000)(8635.000,3012.000)(8395.000,3072.000)
\path(8635,3012)(9235,3012)
\path(8995.000,2952.000)(9235.000,3012.000)(8995.000,3072.000)
\path(8635,3012)(9235,3312)
\path(9047.170,3151.003)(9235.000,3312.000)(8993.505,3258.334)
\path(9235,3012)(9835,3012)
\path(9595.000,2952.000)(9835.000,3012.000)(9595.000,3072.000)
\path(8035,3012)(8635,3312)
\path(8447.170,3151.003)(8635.000,3312.000)(8393.505,3258.334)
\path(8635,2112)(9235,2112)
\path(8995.000,2052.000)(9235.000,2112.000)(8995.000,2172.000)
\path(8635,2112)(9235,1812)
\path(8993.505,1865.666)(9235.000,1812.000)(9047.170,1972.997)
\path(5035,4812)(5635,5112)
\path(5447.170,4951.003)(5635.000,5112.000)(5393.505,5058.334)
\path(5035,4812)(5635,5412)
\path(5507.721,5199.868)(5635.000,5412.000)(5422.868,5284.721)
\put(5713,9008){\ellipse{2996}{2996}}
\put(2113,9008){\ellipse{2996}{2996}}
\put(8413,6608){\ellipse{2996}{2996}}
\put(1513,5708){\ellipse{2996}{2996}}
\put(5335,1512){\ellipse{2996}{2996}}
\put(1813,2258){\ellipse{2996}{2996}}
\put(5113,5408){\ellipse{2996}{2996}}
\put(9235,3312){\whiten\ellipse{212}{212}}
\put(9235,3312){\ellipse{212}{212}}
\put(2035,8412){\blacken\ellipse{212}{212}}
\put(2035,8412){\ellipse{212}{212}}
\put(835,9012){\blacken\ellipse{212}{212}}
\put(835,9012){\ellipse{212}{212}}
\put(1435,9312){\whiten\ellipse{212}{212}}
\put(1435,9312){\ellipse{212}{212}}
\put(2635,9012){\whiten\ellipse{212}{212}}
\put(2635,9012){\ellipse{212}{212}}
\put(3235,9012){\whiten\ellipse{212}{212}}
\put(3235,9012){\ellipse{212}{212}}
\put(2635,9312){\whiten\ellipse{212}{212}}
\put(2635,9312){\ellipse{212}{212}}
\put(3235,9312){\whiten\ellipse{212}{212}}
\put(3235,9312){\ellipse{212}{212}}
\put(2635,8712){\whiten\ellipse{212}{212}}
\put(2635,8712){\ellipse{212}{212}}
\put(3235,8712){\whiten\ellipse{212}{212}}
\put(3235,8712){\ellipse{212}{212}}
\put(3235,8412){\whiten\ellipse{212}{212}}
\put(3235,8412){\ellipse{212}{212}}
\put(2635,9612){\whiten\ellipse{212}{212}}
\put(2635,9612){\ellipse{212}{212}}
\put(2035,9612){\whiten\ellipse{212}{212}}
\put(2035,9612){\ellipse{212}{212}}
\put(2035,9312){\whiten\ellipse{212}{212}}
\put(2035,9312){\ellipse{212}{212}}
\put(2035,9012){\whiten\ellipse{212}{212}}
\put(2035,9012){\ellipse{212}{212}}
\put(835,5412){\whiten\ellipse{212}{212}}
\put(835,5412){\ellipse{212}{212}}
\put(1435,5112){\whiten\ellipse{212}{212}}
\put(1435,5112){\ellipse{212}{212}}
\put(235,5712){\blacken\ellipse{212}{212}}
\put(235,5712){\ellipse{212}{212}}
\put(835,6012){\whiten\ellipse{212}{212}}
\put(835,6012){\ellipse{212}{212}}
\put(2035,5712){\whiten\ellipse{212}{212}}
\put(2035,5712){\ellipse{212}{212}}
\put(2635,5712){\whiten\ellipse{212}{212}}
\put(2635,5712){\ellipse{212}{212}}
\put(2035,6012){\whiten\ellipse{212}{212}}
\put(2035,6012){\ellipse{212}{212}}
\put(2635,6012){\whiten\ellipse{212}{212}}
\put(2635,6012){\ellipse{212}{212}}
\put(2035,5412){\whiten\ellipse{212}{212}}
\put(2035,5412){\ellipse{212}{212}}
\put(2635,5412){\whiten\ellipse{212}{212}}
\put(2635,5412){\ellipse{212}{212}}
\put(2635,5112){\whiten\ellipse{212}{212}}
\put(2635,5112){\ellipse{212}{212}}
\put(2035,5112){\whiten\ellipse{212}{212}}
\put(2035,5112){\ellipse{212}{212}}
\put(2035,4812){\blacken\ellipse{212}{212}}
\put(2035,4812){\ellipse{212}{212}}
\put(2035,6312){\whiten\ellipse{212}{212}}
\put(2035,6312){\ellipse{212}{212}}
\put(1435,6312){\whiten\ellipse{212}{212}}
\put(1435,6312){\ellipse{212}{212}}
\put(1435,6012){\whiten\ellipse{212}{212}}
\put(1435,6012){\ellipse{212}{212}}
\put(1435,5712){\whiten\ellipse{212}{212}}
\put(1435,5712){\ellipse{212}{212}}
\put(1135,1962){\whiten\ellipse{212}{212}}
\put(1135,1962){\ellipse{212}{212}}
\put(1735,1662){\whiten\ellipse{212}{212}}
\put(1735,1662){\ellipse{212}{212}}
\put(535,2262){\whiten\ellipse{212}{212}}
\put(535,2262){\ellipse{212}{212}}
\put(1135,2562){\whiten\ellipse{212}{212}}
\put(1135,2562){\ellipse{212}{212}}
\put(2335,2262){\whiten\ellipse{212}{212}}
\put(2335,2262){\ellipse{212}{212}}
\put(2335,2562){\whiten\ellipse{212}{212}}
\put(2335,2562){\ellipse{212}{212}}
\put(2935,2562){\whiten\ellipse{212}{212}}
\put(2935,2562){\ellipse{212}{212}}
\put(2335,1662){\whiten\ellipse{212}{212}}
\put(2335,1662){\ellipse{212}{212}}
\put(2335,1362){\whiten\ellipse{212}{212}}
\put(2335,1362){\ellipse{212}{212}}
\put(2335,2862){\whiten\ellipse{212}{212}}
\put(2335,2862){\ellipse{212}{212}}
\put(1735,2862){\whiten\ellipse{212}{212}}
\put(1735,2862){\ellipse{212}{212}}
\put(1735,2562){\whiten\ellipse{212}{212}}
\put(1735,2562){\ellipse{212}{212}}
\put(1735,2262){\whiten\ellipse{212}{212}}
\put(1735,2262){\ellipse{212}{212}}
\put(4435,5112){\whiten\ellipse{212}{212}}
\put(4435,5112){\ellipse{212}{212}}
\put(5035,4812){\whiten\ellipse{212}{212}}
\put(5035,4812){\ellipse{212}{212}}
\put(3835,5412){\blacken\ellipse{212}{212}}
\put(3835,5412){\ellipse{212}{212}}
\put(4435,5712){\whiten\ellipse{212}{212}}
\put(4435,5712){\ellipse{212}{212}}
\put(5635,5412){\whiten\ellipse{212}{212}}
\put(5635,5412){\ellipse{212}{212}}
\put(6235,5412){\whiten\ellipse{212}{212}}
\put(6235,5412){\ellipse{212}{212}}
\put(5635,5712){\whiten\ellipse{212}{212}}
\put(5635,5712){\ellipse{212}{212}}
\put(6235,5712){\whiten\ellipse{212}{212}}
\put(6235,5712){\ellipse{212}{212}}
\put(5635,5112){\whiten\ellipse{212}{212}}
\put(5635,5112){\ellipse{212}{212}}
\put(6235,5112){\whiten\ellipse{212}{212}}
\put(6235,5112){\ellipse{212}{212}}
\put(6235,4812){\whiten\ellipse{212}{212}}
\put(6235,4812){\ellipse{212}{212}}
\put(5635,4812){\blacken\ellipse{212}{212}}
\put(5635,4812){\ellipse{212}{212}}
\put(5635,4512){\whiten\ellipse{212}{212}}
\put(5635,4512){\ellipse{212}{212}}
\put(5635,6012){\whiten\ellipse{212}{212}}
\put(5635,6012){\ellipse{212}{212}}
\put(5035,6012){\whiten\ellipse{212}{212}}
\put(5035,6012){\ellipse{212}{212}}
\put(5035,5712){\whiten\ellipse{212}{212}}
\put(5035,5712){\ellipse{212}{212}}
\put(5035,8712){\whiten\ellipse{212}{212}}
\put(5035,8712){\ellipse{212}{212}}
\put(5635,8412){\whiten\ellipse{212}{212}}
\put(5635,8412){\ellipse{212}{212}}
\put(4435,9012){\blacken\ellipse{212}{212}}
\put(4435,9012){\ellipse{212}{212}}
\put(5035,9312){\whiten\ellipse{212}{212}}
\put(5035,9312){\ellipse{212}{212}}
\put(6235,9012){\whiten\ellipse{212}{212}}
\put(6235,9012){\ellipse{212}{212}}
\put(6835,9012){\whiten\ellipse{212}{212}}
\put(6835,9012){\ellipse{212}{212}}
\put(6235,9312){\whiten\ellipse{212}{212}}
\put(6235,9312){\ellipse{212}{212}}
\put(6835,9312){\whiten\ellipse{212}{212}}
\put(6835,9312){\ellipse{212}{212}}
\put(6235,8412){\whiten\ellipse{212}{212}}
\put(6235,8412){\ellipse{212}{212}}
\put(6235,8112){\whiten\ellipse{212}{212}}
\put(6235,8112){\ellipse{212}{212}}
\put(6235,9612){\whiten\ellipse{212}{212}}
\put(6235,9612){\ellipse{212}{212}}
\put(5635,9612){\whiten\ellipse{212}{212}}
\put(5635,9612){\ellipse{212}{212}}
\put(5635,9312){\whiten\ellipse{212}{212}}
\put(5635,9312){\ellipse{212}{212}}
\put(5635,9012){\whiten\ellipse{212}{212}}
\put(5635,9012){\ellipse{212}{212}}
\put(8035,2412){\whiten\ellipse{212}{212}}
\put(8035,2412){\ellipse{212}{212}}
\put(8635,2112){\whiten\ellipse{212}{212}}
\put(8635,2112){\ellipse{212}{212}}
\put(7435,2712){\blacken\ellipse{212}{212}}
\put(7435,2712){\ellipse{212}{212}}
\put(8035,3012){\whiten\ellipse{212}{212}}
\put(8035,3012){\ellipse{212}{212}}
\put(9235,3012){\whiten\ellipse{212}{212}}
\put(9235,3012){\ellipse{212}{212}}
\put(9835,3012){\whiten\ellipse{212}{212}}
\put(9835,3012){\ellipse{212}{212}}
\put(9235,2112){\whiten\ellipse{212}{212}}
\put(9235,2112){\ellipse{212}{212}}
\put(9235,1812){\whiten\ellipse{212}{212}}
\put(9235,1812){\ellipse{212}{212}}
\put(1435,8712){\whiten\ellipse{212}{212}}
\put(1435,8712){\ellipse{212}{212}}
\put(8635,3312){\whiten\ellipse{212}{212}}
\put(8635,3312){\ellipse{212}{212}}
\put(8635,3012){\whiten\ellipse{212}{212}}
\put(8635,3012){\ellipse{212}{212}}
\end{picture}
}
\end{center}

\caption{{\bf Epidemic outbreak on a structured population:} Schematic
representation of a population structured in communities (big circles) and
and the spread of an infectious disease inside and between communities.  
The individuals inside a community are divided between locals (open
circles)  and bridges (filled circles). The locals transmit the disease to
other individuals inside the community (solid arcs) while the bridges
transmit the disease to individuals in other communities (dashed arcs).
For simplicity individuals that are not affected by the outbreak are not
shown.}

\label{fig1}
\end{figure}
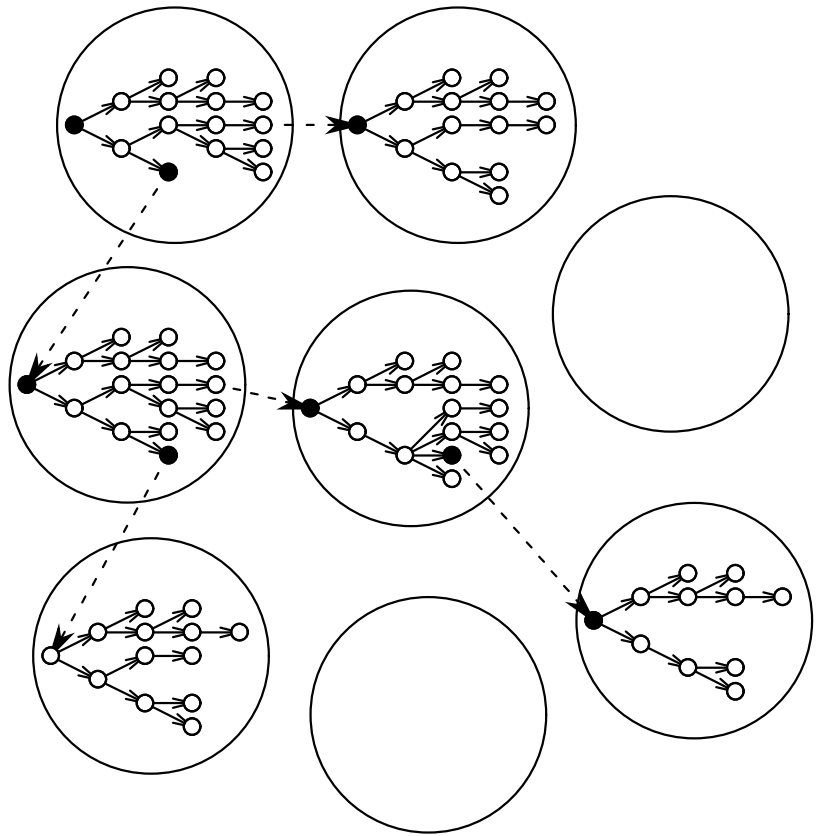

\section*{Model}

Figure \ref{fig1} illustrates the general features of an epidemic outbreak
on a population structured in different communities. Starting from an
index case a disease spreads widely inside a community thanks to the
frequent intra-community interactions. In addition the disease is
transmitted to other communities via individuals belonging to different
communities.  While the inter-community interactions may be rare they are
determinant to understand the overall outbreak progression. Based on this
picture I divide the population in two types or classes. The {\it locals}
belonging to a single community and the social {\it bridges} belonging to
different communities. In a first approximation I assume that (i) all
communities are statistically equivalent, (ii) the mixing between the
local and bridges is homogeneous, and (iii) social bridges belong to two
populations. While these assumptions are off course approximations they
allow us to gain insight into the problem. They could be relaxed in future
works to include other factors such as degree correlations among
interacting individuals \cite{vazquez06e} and more realistic mixing
patterns \cite{vazquez06f}.

An epidemic outbreak taking place inside a community is then modeled by a
a multi-type branching process \cite{mode71} starting from an index case
(see Fig.  \ref{fig1}). The key intra-community magnitudes are the
reproductive number and the generation times \cite{anderson91,vazquez06b}.
The reproductive number is the average number of secondary cases generated
by a primary case. The disease transmission introduces some biases towards
individuals that interact more often. Therefore, I make an explicit
distinction between the index case and other primary cases and denote
their expected reproductive numbers by $R$ and $\tilde{R}$, respectively.  
The generation time $\tau$ is the time elapse from the infection of a
primary case and the infection of a secondary case. It is a random
variable characterized by the generation time distribution function
$G(\tau)$. These magnitudes can be calculated for different models such as
the susceptible infected recovered (SIR) model and they can be estimated
from empirical data as well. Finally, a community outbreak is represented
by a causal true rooted at the index case \cite{vazquez06a,vazquez06b}. In
this tree the generation of an infected case is given by the distance to
the index case. Furthermore, the tree can have at most $D$ generations,
where $D$ is the average distance between individuals inside a community.

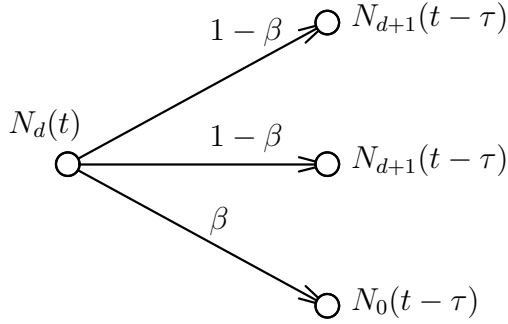
\begin{figure}

\begin{center}
\setlength{\unitlength}{0.00041667in}
\begingroup\makeatletter\ifx\SetFigFont\undefined%
\gdef\SetFigFont#1#2#3#4#5{%
  \reset@font\fontsize{#1}{#2pt}%
  \fontfamily{#3}\fontseries{#4}\fontshape{#5}%
  \selectfont}%
\fi\endgroup%
{\renewcommand{\dashlinestretch}{30}
\begin{picture}(4350,4045)(0,-10)
\thicklines
\path(750,2058)(4050,2058)
\path(3690.000,1968.000)(4050.000,2058.000)(3690.000,2148.000)
\path(750,2058)(4050,258)
\path(3690.861,351.376)(4050.000,258.000)(3777.054,509.397)
\path(750,2058)(4050,3858)
\path(3777.054,3606.603)(4050.000,3858.000)(3690.861,3764.624)
\put(4050,258){\whiten\ellipse{300}{300}}
\put(4050,258){\ellipse{300}{300}}
\put(0,2358){\makebox(0,0)[lb]{{\SetFigFont{12}{14.4}{\familydefault}{\mddefault}{\updefault}$N_d(t)$}}}
\put(2550,3558){\makebox(0,0)[lb]{{\SetFigFont{12}{14.4}{\familydefault}{\mddefault}{\updefault}$1-\beta$}}}
\put(2550,1158){\makebox(0,0)[lb]{{\SetFigFont{12}{14.4}{\familydefault}{\mddefault}{\updefault}$\beta$}}}
\put(4350,3708){\makebox(0,0)[lb]{{\SetFigFont{12}{14.4}{\familydefault}{\mddefault}{\updefault}$N_{d+1}(t-\tau)$}}}
\put(4350,1908){\makebox(0,0)[lb]{{\SetFigFont{12}{14.4}{\familydefault}{\mddefault}{\updefault}$N_{d+1}(t-\tau)$}}}
\put(2550,2208){\makebox(0,0)[lb]{{\SetFigFont{12}{14.4}{\familydefault}{\mddefault}{\updefault}$1-\beta$}}}
\put(750,2058){\whiten\ellipse{300}{300}}
\put(750,2058){\ellipse{300}{300}}
\put(4350,108){\makebox(0,0)[lb]{{\SetFigFont{12}{14.4}{\familydefault}{\mddefault}{\updefault}$N_0(t-\tau)$}}}
\put(4050,2058){\whiten\ellipse{300}{300}}
\put(4050,2058){\ellipse{300}{300}}
\put(4050,3858){\whiten\ellipse{300}{300}}
\put(4050,3858){\ellipse{300}{300}}
\end{picture}
}
\end{center}

\caption{{\bf Local disease transmission:} Diagram representing the
disease transmission from a primary case at generation $d$ to secondary
cases in the following generation. The secondary cases are locals with
probability $1-\beta$, potentially leading to subsequent infections inside
their community, and bridges with probability $\beta$, transmitting the
disease to other communities. Note that the expected number of 
descendants generated by a secondary case is evaluated at a delayed time 
$t-\tau$, where $\tau$ is the generation time.}

\label{fig2}
\end{figure}

\section*{Spreading regimes}

Let us focus on a primary case at generation $d$ and its secondary cases 
at the following generation (see Fig. \ref{fig2}). Let $N_d(t)$ denote the 
expected number of descendants of the primary case at generation $d$. In 
particular $N_0(t)$ gives the expected number of descendants from the 
index case, i.e. the expected outbreak size. In turn, $N_{d+1}(t)$ is the 
expected number of descendants generated by a local secondary case at 
generation $d+1$. Otherwise, if the secondary case is a bridge, it starts 
a new outbreak in a different community with expected outbreak size 
$N_0(t)$. Putting together the contribution of locals and bridges we 
obtain the recursive equation

\begin{equation}
N_d(t) = \left\{
\begin{array}{ll}
\displaystyle
(1-\beta)\left[ 1 + R \int_0^tdG(\tau) N_{d+1}(t-\tau) \right]\ , & d=0\\
\\
(1-\beta)\left[ 1 + \tilde{R} \int_0^tdG(\tau) N_{d+1}(t-\tau)\right]
+\beta N_0(t)\ , & 0<d<D\\
\\
1-\beta +\beta N_0(t)\ , & d=D\ .
\end{array}
\right.
\label{NN}
\end{equation}

\noindent
Iterating this equation from $d=D$ to $d=0$ we obtain

\begin{equation}
N_0(t) = 1 + (1-\beta) F(t) + \beta \int_0^tdF(\tau)N_0(t-\tau)\ ,
\label{N0N0}
\end{equation}

\noindent where

\begin{equation}
F(t) = R \sum_{d=1}^D \left[ (1-\beta)\tilde{R} \right]^{d-1} 
G^{\star d}(t)
\label{Ft}
\end{equation}

\noindent and $G^{\star d}(t)$ denotes the $d$-order convolution of
$G(t)$, i.e. $G^{\star 0}(t)=1$ and $G^{\star d+1}(t) =
\int_0^tdG(\tau)G^{\star d}(t-\tau)$. $F(t)$ represents the expected
outbreak size inside a community at time $t$ and 

\begin{equation}
N_{\rm C} = \lim_{t\rightarrow\infty} F(t)\ ,
\label{nNoo}
\end{equation}

\noindent is the final expected outbreak size inside a community. When
$\beta=0$ it coincides with the expected outbreak size inside a community
\cite{vazquez06b}. When $\beta>0$ (\ref{N0N0}) provides a
self-consistent equation to determine the overall expected outbreak size
after taking into account the inter-community transmissions.

To calculate $N_0(t)$ I use the Laplace transform method. Consider the
incidence

\begin{equation}
I(t) = \dot{N}_0(t)
\label{nN}
\end{equation}

\noindent and its Laplace transform

\begin{equation}
\hat{I}(\omega) = \int_0^\infty dt e^{-\omega t} I(t)\ .
\label{nomega}
\end{equation}

\noindent Substituting the recursive equation (\ref{N0N0}) in
(\ref{nomega}) I obtain

\begin{equation}
\hat{I}(\omega) =
\frac{\hat{f}(\omega)}{1-\beta \hat{f}(\omega)}\ ,
\label{fomega1}
\end{equation}

\noindent where

\begin{equation}
\hat{f}(\omega) = \int_0^\infty dt e^{-\omega t} \dot{F}(t)\ .
\label{fomega}
\end{equation}

\noindent The validity of (\ref{nomega}) is restricted to $\omega$ values 
satisfying $1-\beta \hat{f}(\omega)>0$, resulting in different scenarios 
depending on the value of the parameter

\begin{equation}
R_{\rm C} = \beta N_{\rm C}\ .
\label{P0}
\end{equation}

\noindent {\it Local outbreaks:} When $R_{\rm C}<1$ then $\hat{I}(\omega)$
is defined for all $\omega\geq0$ and $I(t)$ is obtained inverting the
Laplace transform in (\ref{nomega}).  Furthermore, since $\hat{I}(0)$ is
defined from (\ref{fomega1}) it follows that $I(t)$ decreases to zero when
$t\rightarrow\infty$, i.e. the epidemic outbreak dies out.

\noindent {\it Global outbreaks:} When $R_{\rm C}>1$ the incidence grows
exponentially $I(t)\sim e^{\omega_c t}$, where $\omega_c$ is the positive
root of the equation

\begin{equation}
\beta\hat{f}(\omega)=1\ .
\label{omegac}
\end{equation}

\noindent These two scenarios are equivalent to those obtained for a
single community \cite{anderson91}. $R_{\rm C}$ represents the effective
community's reproductive number and the threshold condition

\begin{equation}
R_{\rm C} = 1
\label{betac}
\end{equation}

\noindent delimits the local and global scenarios.

To go beyond the final outbreak I analyze the progression of the
inter-communities outbreak. I assume that the disease is transmitted at a
constant rate $\lambda$ from a primary case to a secondary case
independently of their type. In this case the intra-community incidence is
given by \cite{vazquez06b}

\begin{equation}
\dot{F}(t) \sim N_{\rm C} \frac{\lambda (\lambda t)^{D-1} 
e^{-\lambda t}}{(D-1)!}\ ,
\label{ftgamma}
\end{equation}

\noindent for $t\gg t_0$, where

\begin{equation}
t_0= \frac{D-1}{\tilde{R}} \frac{1}{\lambda}\ .
\label{t0}
\end{equation}

\noindent Calculating the inverse Laplace transform of
(\ref{nomega}) I finally obtain

\begin{equation}
I(t) = N_{\rm C} \sum_{m=0}^\infty (\beta N_\infty)^m
\frac{\lambda (\lambda t)^{D(m+1)-1} 
e^{-\lambda t}}{\Gamma[D(m+1)]}\ ,
\label{ntgamma}
\end{equation}

\begin{figure}

\centerline{\includegraphics[width=3.5in]{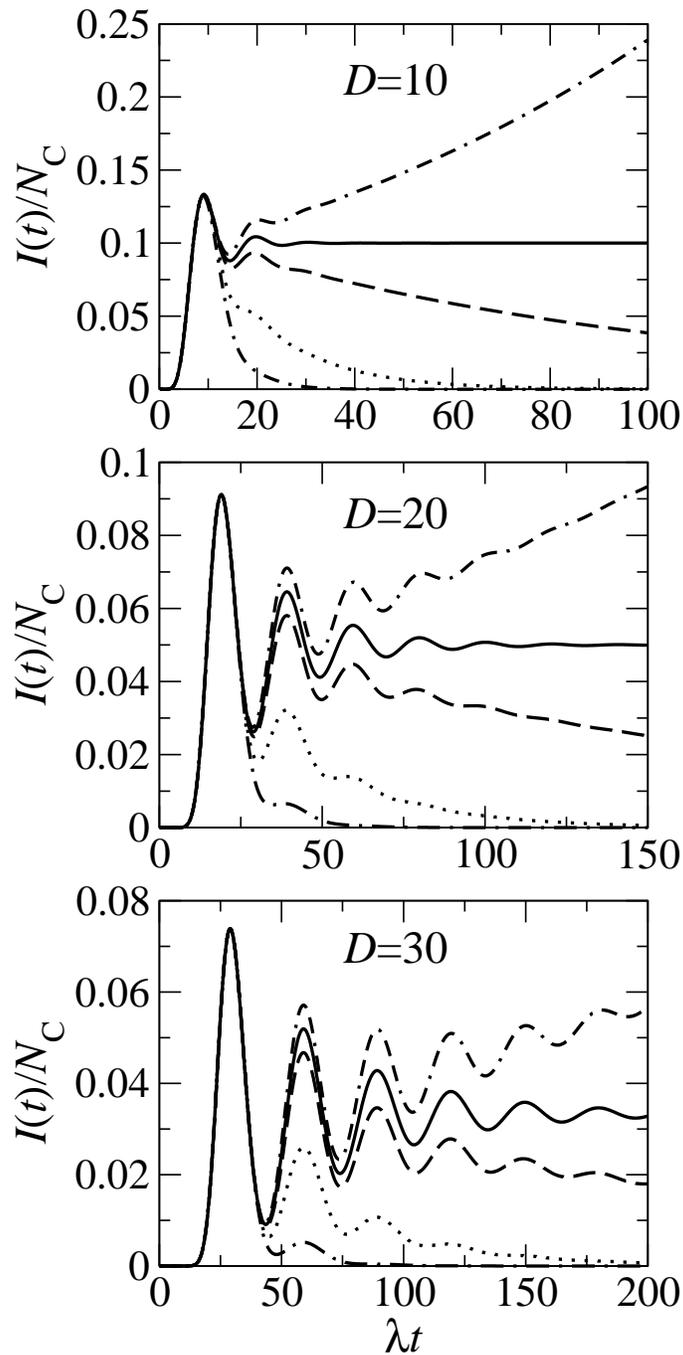}}

\caption{{\bf Epidemic outbreak progression:} The incidence $I(t)$ as a
function of time in units of the local disease transmission rate
$\lambda$, for $R_{\rm C}=0.1$ (dash-dotted), 0.5 (dotted), 0.9
(dashed), 1.0 (solid) and 1.1 (dash-dash-dotted). The panels from top to
bottom corresponds to different average distances $D$ between individuals
inside a community.}

\label{fig4}
\end{figure}

\noindent where $\Gamma(x)$ is the gamma function. Figure \ref{fig4} shows
the progression of the incidence as obtained from (\ref{ntgamma}). As
predicted above, the outbreak dies out when $R_{\rm C}<1$ while when $R_{\rm C}>1$ it
grows exponentially. More important, the incidence exhibits oscillations
at the early stages, their number increasing with increasing $D$. For
example, we distinguish about two oscillations for $D=10$ while for $D=30$
several oscillations are observed. These oscillations represent resurgent
epidemics, which are often observed in real outbreaks
\cite{riley03,anderson04} and simulations \cite{sattenspiel95,watts05}.

\begin{figure}
\centerline{\includegraphics[width=6in]{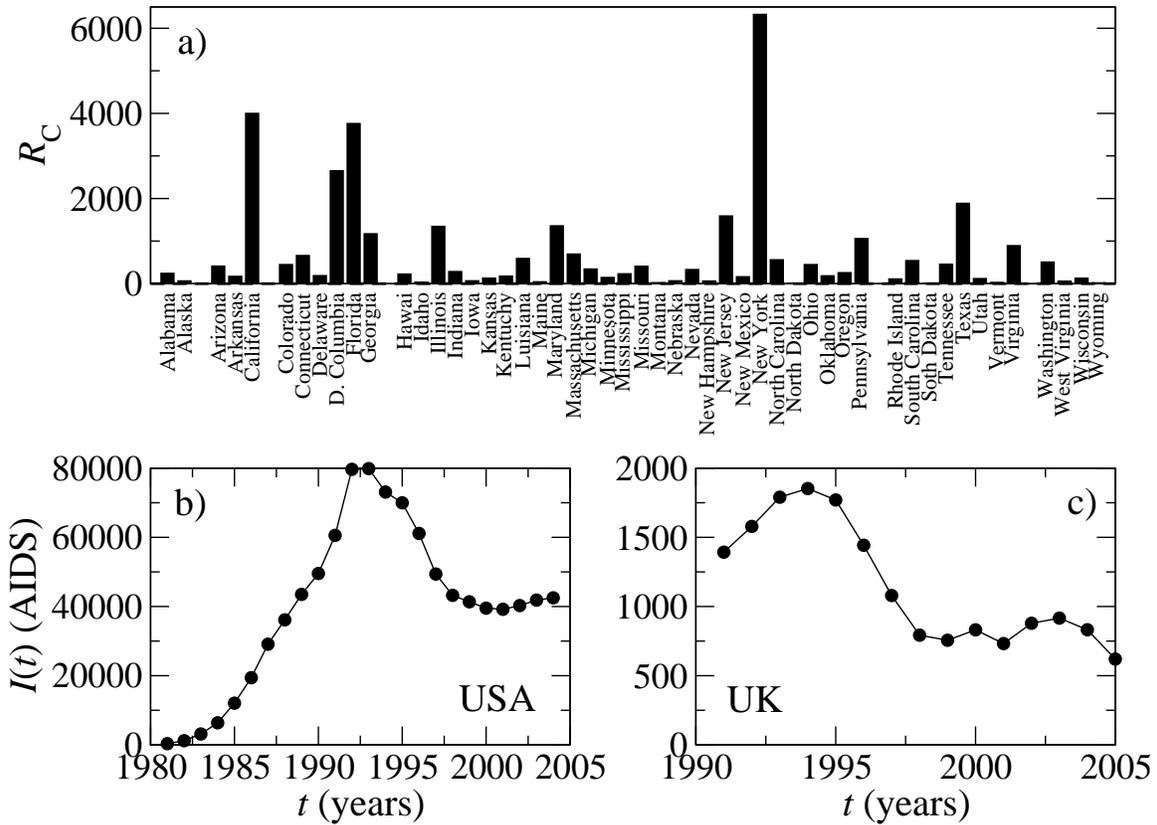}}

\caption{{\bf USA AIDS epidemics:} Estimated $R_{\rm C}=\beta N_{\rm C}$ 
for the
different USA states. $\beta$ was computed as the ratio between the number
of state out-immigrants and the total state population according to the
1995-2000 USA census (http://www.census.gov). $N_{\rm C}$ was computed as
the number of habitants living with AIDS according to the 2005 statistics
published by the US Department of Health (http://www.hhs.gov).  b) and c)
AIDS incidence in the USA b) and UK c) by year, as reported by the US
Department of Health and the UK Health Protection Agency
(http://www.hpa.org.uk), respectively.}

\label{fig3}
\end{figure}

\section*{Case study: AIDS epidemics}

To understand the relevance of these results in a real world scenario I
analyze data reported for the AIDS epidemics. First, I estimate the
parameter $R_{\rm C}$ determining the spreading regime, local or global.
Figure \ref{fig3}a shows the value of $R_{\rm C}$ across the USA by state.
For most states $R_{\rm C}>1$, reaching significantly large values for
several states. For example, $R_{\rm C}$ exceeds 1,000 for California and
New York. These numbers indicate that the USA AIDS epidemics is in the
global spread scenario ($R_{\rm C}>1$), in agreement with the general
believe.

Second, I analyze the temporal evolution of the AIDS incidence. Fig.
\ref{fig3}b and c show the AIDS incidence in USA and UK by year,
indicating a similar temporal pattern. The epidemics started with an 
increasing tendency of the incidence which, after reaching a maximum, 
switched to a decreasing trend. After some years, however, the epidemics 
resurges with a new incidence increase.  This picture coincides with the model 
predictions in Fig.  \ref{fig4}. Therefore, a possible explanation of the 
observed multiple peaks is the existence of a community structure, which 
can be attributed to geographical location and other factors.

\section*{Discussion and conclusions}

$R_{\rm C}$ in (\ref{betac}) represents the expected number of infected
individuals leaving their community. The numerical simulations reported in
\cite{watts05} indicated the existence of a transition at $R_{\rm C}=1$,
from local outbreaks when $R_{\rm C}<1$ to global epidemics when $R_{\rm
C}>1$. I have demonstrated that there is indeed a phase transition at
$R_{\rm C}=1$. Furthermore, the analytical solution provides an expression
of $R_{\rm C}$ as a function of the bridge's fraction and the
intra-community expected outbreak size (\ref{P0}). $R_{\rm C}$ represents
a measure of the reproductive number at the inter-community level. Its
value can be estimated from the expected outbreak size inside a community
and the bridge's fraction. Based on the resulting estimate we can
determine if an epidemics is in the local or global epidemics scenario and
react accordingly.

The inter-community disease transmission is characterized by oscillations
at the early stages which represents resurgent epidemics, the number of
these resurgencies being determined by the characteristic distance between
individuals within a community. In essence, when $D$ is small the time
scale characterizing the outbreak progression within a community is very
small \cite{bbv04,bbv05,vazquez06b}.  Therefore, the time it takes to
observe the infection of a social bridge is very small as well, resulting
in the mixing between the intra- and inter-community transmissions. In
contrast, when $D$ is large it takes a longer time to observe the
infection of a social bridge and by that time the intra-community outbreak
has significantly developed.  Therefore, in this last case the outbreak
within communities is partially segregated in time.

When multi-agent models are not available these results allow us to 
evaluate the potential progression of an epidemic outbreak and 
consequently determine the magnitude of our response to halt it. They are 
also valuable when a detailed metapolpulation model is available, 
funneling the search for key quantities among the several model 
parameters. More important, this work open avenues for future analytical 
works that side by side with multi-agent models will increase our chances 
to control global epidemics.



\end{document}